\documentclass[prd,eqsecnum,floatfix]{revtex4}%
\usepackage{epsfig}
\usepackage{amsmath}
\usepackage{graphicx}
\usepackage{pstricks}%
\usepackage{amsfonts}%
\usepackage{amssymb}
\newtheorem{theorem}{Theorem}
\newtheorem{acknowledgement}[theorem]{Acknowledgement}

\newtheorem{definition}[theorem]{Definition}

\newcommand{\BOX}{\hbox {$\sqcap$ \kern -1em $\sqcup$}}

\newcommand{\be}{\begin{equation}}
\newcommand{\ee}{\end{equation}}
\newcommand{\ba}{\begin{eqnarray}}
\newcommand{\ea}{\end{eqnarray}}
\newcommand{\ban}{\begin{eqnarray*}}
\newcommand{\bea}{\begin{eqnarray}}
\newcommand{\eea}{\end{eqnarray}}
\newcommand{\ean}{\end{eqnarray*}}
\newcommand{\barr}{\begin{array}}
\newcommand{\earr}{\end{array}}

\begin{document}

\title{ Methods and Models for the Study of Decoherence }
\author{ Sarben~Sarkar}
\affiliation{King's College London, University of London, Department of Physics, Strand
WC2R 2LS, London, U.K.}

\begin{abstract}
{\small We shall review methods used in the description of decoherence on particle
probes in experiments due to surrounding media. This will include
conventional media as well as a model for space-time foam arising from
non-critical string theory.}
 \end{abstract}
\maketitle


\section{The role of decoherence}

Until recently in elementary particle physics the environment was not
considered. Scatterings were calculated in a vacuum background and S-matrix
elements were calculated within the paradigm of the standard gauge theory
model. The latter is a successful theory overall. However, recently systems
which oscillate coherently have been investigated with increasing precision,
e.g. neutrino and neutral meson flavour oscillations. Clearly neutrinos
produced in the sun, on going through it, encounter an obvious scattering
environment. In laboratory experiments however there does not seem to be the
need for such considerations; of course there are uncertainties in
determining time and position which lead to features akin to decoherence %
\cite{olsson}. However, triggered again by increased precision, the effect
of fluctuations in the space-time metric due to space-time defects such as
microscopic black holes, and D branes in string theory are being estimated.
Given the smallness of the gravitational coupling compared to the other
interactions in the past the search for such effects was regarded as
optimisitc. Progress in experimental techniques is making such effects more
testable \cite{expts}.

In this Handbook it was considered to be desirable to split the discussion
of decoherence between two chapters.This one will render a brief account of
the methods of decoherence that are used in the analysis of experiments
given in the companion chapter \cite{bernabeu1}. We shall demonstrate why
there is a large universality class in the space of theories describing
decoherence with most analyses using models from this class. However we
should stress that the universality is for descriptions where the
system-environment interaction is in some sense conventional. Indeed when we
introduce descriptions emanating from string theory we can and do produce
descriptions which can give qualitatively different effects \cite{bernabeu}.
Such non-conventional descriptions are to be expected since it is natural
for quantum space-time to be somewhat different from the paradigm of
Brownian phenomena in condensed matter. Moreover the manifestation of
gravitational decoherence in a theory, which is diffeomorphic covariant at
the classical level, is not just restricted to fluctuation and dissipation.
It is pivotal in the breakdown of discrete symmetries such as CPT and more
obviously T. This is an exciting role for decoherence because it gives rise
to qualitatively new phenomena \cite{barenboim} which is being tested now
and in the next generation of laboratory experiments.

\bigskip

This paper will be divided into three sections:

\begin{itemize}
\item decoherence in a general setting with a discussion of how coherence is
lost and the implication for discrete symmetries

\item generic treatment of system-reservoir interactions and the Lindblad
formalism from Markovian approximations

\item non-critical string theory and D-particle foam and the phenomenolgy of
stochastic metrics
\end{itemize}

\section{General Features of Decoherence}

The fact that an environment ${\cal E}$ \ interacts with a system ${\cal S}$
and is affected by \ it is obvious whether they interact classically or
quantum mechanically. However classically the measurement of ${\cal E}$ can
only locally affect ${\cal S}$. This is in sharp contrast to the quantum
mechanical situation where non-local effects can take place. The associated
distinguishing property is \ that of entanglement. For the compound system $%
{\cal ES}$ Schmidt bases allow us to write the state $\left| \Psi
\right\rangle $ as 
\[
\left| \Psi \right\rangle =\sum_{n}\sqrt{p_{n}}\left| \phi _{n}\right\rangle
\left| \Phi _{n}\right\rangle 
\]%
where the Hilbert space $H_{{\cal S}}$ of states $\left| \phi
_{n}\right\rangle $ are associated with ${\cal S}$ and the Hilbert space $H_{%
{\cal E}}$ of states $\left| \Phi _{n}\right\rangle $are associated with with ${\cal E}$. In the Schmidt basis the
states for different $n$ in the different spaces have to be mutually
orthogonal i.e. 
\[
\left\langle \phi _{n}\right| \left. \phi _{m}\right\rangle =\left\langle
\Phi _{n}\right| \left. \Phi _{m}\right\rangle =\delta _{nm}
\]%
and the non-negative coefficients $p_{n}$ satisfy $\sum_{n}p_{n}^{2}=1$.

The corresponding density matrix is%
\[
\rho =\rho _{class.}+\sum_{n\neq m}\sqrt{p_{n}p_{m}}\left| \phi
_{n}\right\rangle \left\langle \phi _{m}\right| \otimes \left| \Phi
_{n}\right\rangle \left\langle \Phi _{m}\right| 
\]%
where $\rho _{class.}=\sum_{n}p_{n}\left| \phi _{n}\right\rangle
\left\langle \phi _{n}\right| \otimes \left| \Phi _{n}\right\rangle
\left\langle \Phi _{n}\right| $. The term $\rho -\rho _{class.}$ is known as
the entanglement. Clearly entanglement is a measure of the departure of the
compound system from a product state of states of ${\cal S}$ and ${\cal E}$.
A classic example of a pure entangled state is the EPR\ state ( Einstein-Podolsky-Rosen) written conventionally in
terms of spin $\frac{1}{2}$ systems 
\[
\frac{\left| \uparrow \right\rangle \left| \downarrow \right\rangle -\left|
\downarrow \right\rangle \left| \uparrow \right\rangle }{\sqrt{2}} 
\]%
which is clearly not factorisable. Now let us see how the interaction between ${\cal %
S}$ and ${\cal E}$ leads to decoherence\ by considering a simple interaction 
\[
\lambda H_{{\cal ES}}=\sum_{n}\left| \phi _{n}\right\rangle \left\langle
\phi _{n}\right| \otimes \widehat{A}_{n} 
\]%
where $\widehat{A}_{n}$ are operators on the $H_{{\cal E}}$. For an initial
pure unentangled state i.e. a product state 
\[
\left| \Psi \right\rangle =\sum_{n}c_{n}\left| \phi _{n}\right\rangle \left|
\Theta _{0}\right\rangle 
\]

(where $\left| \Theta _{0}\right\rangle $ can be expressed in terms of \ the 
$\left| \Phi _{n}\right\rangle $s) under time evolution 
\[
\left| \phi _{n}\right\rangle \left| \Theta _{0}\right\rangle \stackrel{t}{%
\longrightarrow }\left| \phi _{n}\right\rangle \exp \left( -i\widehat{A}%
_{n}t\right) \left| \Theta _{0}\right\rangle \equiv \left| \phi
_{n}\right\rangle \left| \Theta _{n}\left( t\right) \right\rangle 
\]

\bigskip\ The density matrix traced over the environment $\rho _{{\cal S}%
}\left( t\right) $ gives 
\[
\rho _{{\cal S}}\left( t\right) =\sum_{n,m}c_{m}^{\ast }c_{n}\left\langle
\Theta _{m}\left( t\right) \right. \left| \Theta _{n}\left( t\right)
\right\rangle \left| \phi _{m}\right\rangle \left\langle \phi _{n}\right| 
\]

If the circumstances are such that $\left\langle \Theta _{m}\left( t\right)
\right. \left| \Theta _{n}\left( t\right) \right\rangle \longrightarrow
\delta _{mn}$ as $t\longrightarrow \infty $, \ then asymptotically 
\[
\rho _{{\cal S}}\left( t\right) \longrightarrow \sum_{n}\left| c_{n}\right|
^{2}\left| \phi _{n}\right\rangle \left\langle \phi _{n}\right| .
\]%
All coherences embodied by off-diagonal matrix elements have vanished, i.e. there is complete decoherence \cite{giulini}.

\bigskip

We will now consider an {\it associated} aspect of the interaction of the
system with the environment, the lack of an invertible scattering matrix.
Consider schematically three spaces ${\frak H}_{1},{\frak H}_{2}$ and $%
{\frak H}_{3}$ where ${\frak H}_{1}$ is the space of states of the initial
states, ${\frak H}_{2}$ is the state space for inaccessible environmental
degrees of freedom (e.g. states inside a black hole horizon) and ${\frak H}_{3}$ is the space of final states. Within
a scattering matrix formalism consider an in-state $\sum_{A}x_{A}\left|
X_{A}\right\rangle _{1}\left| 0\right\rangle _{2}\left| 0\right\rangle _{3}$
(where the subscripts \ $1$, $2$ and $3$ are related to the spaces ${\frak H}%
_{1},{\frak H}_{2}$ and ${\frak H}_{3}$) this is scattered to $\sum_{A}{\Bbb %
S}_{A}^{\;bc}x_{A}\left| 0\right\rangle _{1}\left| \overline{Y}%
^{b}\right\rangle _{2}\left| \overline{Z}^{c}\right\rangle _{3}$ where \ the
bar above the state labels indicates the CPT\ transform \cite{page}. (\ On
introducing the operator $\theta =CPT$ we have explicitly $\left| \overline{Y%
}^{b}\right\rangle =\theta \left| Y_{b}\right\rangle $ etc.) Now on tracing
over the inaccessible degrees of freedom ( in ${\frak H}_{2}$ ) we obtain 
\[
\left| X_{A}\right\rangle \left\langle X_{A}\right| \longrightarrow
\sum_{c,c^{\prime }}\not{S}_{A\;A}^{\;c\;\,\,c^{\prime }}\left| \overline{Z}%
^{c}\right\rangle \left\langle \overline{Z}^{c^{\prime }}\right| 
\]%
with the effective scattering matrix $\not{S}$ given by 
\[
\not{S}_{A\;A}^{\;c\;\,\,c^{\prime }}=\sum_{b,b^{\prime }}{\Bbb S}_{A}^{\;bc}%
{\Bbb S}_{A}^{\ast \;b^{\prime }c^{\prime }}.
\]%
This does not factorise, which it would have to, for $\not{S}$ to be of the
form $UU^{\dagger }$. \ \ Consequently evolution is non-unitary. This is
generic to environmental decoherence. \ Of course with space-time defects
the inaccessible degrees of freedom can be behind causal horizons.

\bigskip

For local relativistic \ interacting quantum field theories there is the
CPT\ theorem. Such theories show unitary evolution. A violation of CPT\ for
Wightman functions ( i.e. unordered correlation functions for fields)\
implies violation of Lorentz invariance \cite{greenberg}. However CPT\
invariance of course is not sufficient for Lorentz invarilance. For physical
systems, which in the absence of gravity show CPT\ invariance, the
incorporation of a gravitational environment can lead to non-unitary
evolution as we have argued. In fact we shall sketch arguments from
non-critical string theory which produce such non-unitary evolution. There
is then a powerful argument due to Wald which argues that an operator $%
\theta $ incorporating strong CPT\ invariance does not exist. The argument
proceeds via reduction ad absurdum. For strong CPT\ invariance to hold we
should have in states and out states connected by $\not{S}$ and $\theta $
and their operations commute in the following sense. For an in state $\rho
_{in}$ there is an out state $\rho _{out}$ such that 
\[
\rho _{out}=\not{S}\rho _{in}.
\]%
Also there is another out state $\rho _{out}^{\prime }=\theta \rho _{in}$
associated with $\rho _{in}$. If the CPT transforms of states have the same $%
\not{S}$ evolution as the untransformed states then there is strong CPT
invariance. In such situations 
\[
\theta \not{S}\theta \not{S}\rho _{in}=\rho _{in}
\]%
and so 
\[
\theta \not{S}\theta \not{S}=I,
\]

\bigskip i.e. $\not{S}$ has an inverse. In most circumstances interaction
with an environment produces dissipation and so the inverse of $\not{S}$
would not exist. Hence the assumption of strong CPT is incompatible with
non-unitary evolution \cite{wald}.

\bigskip\ 

\section{Particles propagating in a medium and master equations}

\bigskip

Particles reaching us from outside a laboratory always travel through some
physical medium which can often be described by a conventional medium. For
the moment we will be general and call the medium ${\cal E}$ and the
particle ${\cal S}$. \ We are ignoring particle-particle interactions and so
the approximation of a single body point of view is appropriate. This
bipartite separation can be subtle since different degrees of freedom of the
same particle can be distributed between ${\cal E}$ and ${\cal S}$.
Initially (at $t=t_{0}$) the state $\rho $ of the compound system is assumed
to have a factorised form%
\begin{equation}
\rho \left( t_{0}\right) =\rho _{{\cal S}}\otimes \rho _{{\cal E}}
\label{separability}
\end{equation}%
with $\rho _{{\cal S}}$ being a normalised density operator on the Hilbert
space ${\frak H}_{{\cal S}}$ of states of ${\cal S}$ and analogously for $%
\rho _{{\cal E}}$ . This condition may be not hold in the very early
universe and for an ever present meidum such as space-time foam; ${\cal E}$
and ${\cal S}$ would then always be entangled. Certainly for laboratory
experiments the condition \ref{separability} is acceptable \cite{pike} and the analysis is simplified.
Write the total hamiltonian $H$ as 
\[
H=H_{{\cal S}}+H_{{\cal E}}+\widehat{H}_{S{\cal E}}
\]%
where $H_{S{\cal E}}$ represents the interaction coupling the system and
environment.The Heisenberg equation is%
\begin{equation}
\frac{\partial \rho }{\partial t}=-i\left[ H_{{\cal S}}+H_{{\cal E}}+H_{S%
{\cal E}},\rho \right] \equiv L\rho   \label{liouville}
\end{equation}%
and we will also find it useful to let $-i\left[ H_{{\cal S}},\rho \right]
\equiv L_{{\cal S}}\rho ,\;-i\left[ H_{{\cal E}},\rho \right] \equiv L_{%
{\cal E}}\rho $ and $-i\left[ H_{{\cal SE}},\rho \right] \equiv L_{{\cal SE}%
}\rho $. $\rho $ evolves unitarily. For measuring with operators acting on $%
{\frak H}_{{\cal S}}$ it is sufficient to consider 
\begin{equation}
\rho _{{\cal S}}=Tr_{B}\rho 
\end{equation}%
but given a $\rho _{{\cal S}}$ there is in general no unique $\rho $ associated with it. 
Hence the evoultion of $\rho _{{\cal S}}$ is not well defined. However by
choosing a reference environment state $\rho _{{\cal E}}$ satisfying 
\begin{equation}
L_{{\cal E}}\rho _{{\cal E}}=0  \label{env}
\end{equation}%
we can associate with a $\rho _{{\cal S}}$ a unique state $\rho _{{\cal S}%
}\otimes \rho _{{\cal E}}$ of $S{\cal E}$. In this way a well defined
evolution can be envisaged.

\bigskip

We will obtain a master equation for $\rho _{{\cal S}}$ \ by using the
method of projectors \cite{nakajima}. Let us define 
\[
P\rho =\left( Tr_{{\cal E}}\rho \right) \otimes \rho _{{\cal E}}.
\]%
Clearly 
\[
P^{2}\rho =\left[ Tr_{{\cal E}}\rho _{{\cal E}}\right] \left( Tr_{{\cal E}%
}\rho \right) \otimes \rho _{{\cal E}}=\left( Tr_{{\cal E}}\rho \right)
\otimes \rho _{{\cal E}}=P\rho 
\]%
and so $P$ is a projector. Also we define $Q=1-P$. Acting on \ref{liouville}
with $P$ gives 
\begin{equation}
P\frac{\partial \rho }{\partial t}=PL\rho =PLP\rho +PLQ\rho .  \label{proj1}
\end{equation}%
Similarly 
\begin{equation}
Q\frac{\partial \rho }{\partial t}=QL\rho =QLP\rho +QLQ\rho .  \label{proj2}
\end{equation}%
These give two coupled equations for $P\rho $ and $Q\rho .$ \ref{proj2} can
be solved for $Q\rho $ on noting that 
\[
\left( \frac{\partial }{\partial t}-QL\right) Q\rho =QLP\rho 
\]%
and then on formally integrating 
\[
\int_{0}^{t}\frac{\partial }{\partial t^{\prime }}\left( e^{-QLt^{\prime
}}Q\rho \left( t^{\prime }\right) \right) dt^{\prime
}=\int_{0}^{t}e^{-QLt^{\prime }}QLP\rho \left( t^{\prime }\right) dt^{\prime
}
\]%
i.e. 
\begin{equation}
e^{-QLt}Q\rho \left( t\right) =Q\rho \left( 0\right)
+\int_{0}^{t}e^{-QLt^{\prime }}QLP\rho \left( t^{\prime }\right) dt^{\prime
}.
\end{equation}%
This expression for $Q\rho $ is substituted in \ref{proj1} to give 
\[
Tr_{{\cal E}}\left[ P\frac{\partial \rho }{\partial t}\right] =\frac{%
\partial \rho _{{\cal S}}}{\partial t}=Tr_{{\cal E}}\left[ PLP\rho \right]
+Tr_{{\cal E}}\left[ PL\left( e^{-QLt}Q\rho \left( 0\right)
+\int_{0}^{t}e^{-QL\left( t-t^{\prime }\right) }QLP\rho \left( t^{\prime
}\right) dt^{\prime }\right) \right] .
\]%
and can be simplified further on noting that 
\begin{equation}
PL_{{\cal E}}Q\rho =PL_{{\cal E}}\left( \rho -P\rho \right) =PL_{{\cal E}%
}\rho =0\Longrightarrow PL_{{\cal E}}=0  \label{identity1}
\end{equation}%
owing to the cyclic properties of traces. Also 
\begin{equation}
PL_{{\cal S}}Q\rho =PL_{{\cal S}}\left( \rho -\rho _{{\cal S}}\otimes \rho _{%
{\cal E}}\right) =L_{{\cal S}}P\rho -\left( L_{{\cal S}}\rho _{{\cal S}%
}\right) \otimes \rho _{{\cal E}}=0\Longrightarrow PL_{{\cal S}}=PL_{{\cal S}%
}P.  \label{identity2}
\end{equation}%
Hence 
\begin{eqnarray*}
Tr_{{\cal E}}\left( PL_{{\cal S}}P\rho \right)  &=&Tr_{{\cal E}}\left( PL_{%
{\cal S}}\rho \right)  \\
&=&Tr_{{\cal E}}\left( Tr_{{\cal E}}\left( L_{{\cal S}}\rho \right) \otimes
\rho _{{\cal E}}\right) =Tr_{{\cal E}}\left( L_{{\cal S}}\rho \right) =L_{%
{\cal S}}\rho _{{\cal S}}.
\end{eqnarray*}%
Also we assume that $H_{{\cal SE}}=V_{{\cal S}}\otimes V_{{\cal E}}$ (which
is standard for local quantum field theory) and so 
\begin{eqnarray*}
Tr_{{\cal E}}\left( PL_{{\cal SE}}P\rho \right)  &=&Tr_{{\cal E}}\left( PL_{%
{\cal SE}}\rho _{{\cal S}}\otimes \rho _{{\cal E}}\right)  \\
&=&Tr_{{\cal E}}\left[ P\left( V_{{\cal S}}\rho _{{\cal S}}\right) \otimes
\left( V_{{\cal E}}\rho _{{\cal E}}\right) -P\left( \rho _{{\cal S}}V_{{\cal %
S}}\right) \otimes \left( \rho _{{\cal E}}V_{{\cal E}}\right) \right]  \\
&=&Tr_{{\cal E}}\left[ V_{{\cal S}}\rho _{{\cal S}}\otimes \rho _{{\cal E}%
}Tr_{{\cal E}}\left( V_{{\cal E}}\rho _{{\cal E}}\right) -\rho _{{\cal S}}V_{%
{\cal S}}\otimes \rho _{{\cal E}}Tr_{{\cal E}}\left( \rho _{{\cal E}}V_{%
{\cal E}}\right) \right]  \\
&=&\left[ V_{{\cal S}},\rho _{{\cal S}}\right] \otimes \rho _{{\cal E}%
}\left( Tr_{{\cal E}}\left( V_{{\cal E}}\rho _{{\cal E}}\right) \right)  \\
&=&Tr_{{\cal E}}\left( L_{{\cal SE}}\rho _{{\cal E}}\right) \rho _{{\cal S}}.
\end{eqnarray*}%
The analysis would go through also when $H_{{\cal SE}}$ is a sum of
factorised terms. Similarly on using \ref{identity1} and \ref{identity2} 
\[
Tr_{{\cal E}}\left( PLe^{QLt}Q\rho \left( 0\right) \right) =Tr_{{\cal E}%
}\left( L_{{\cal SE}}e^{QLt}Q\rho \left( 0\right) \right) 
\]%
and 
\[
Tr_{{\cal E}}\left( PLe^{QLt^{\prime }}QLP\rho \left( t-t^{\prime }\right)
\right) =Tr_{{\cal E}}\left( L_{{\cal SE}}e^{QLt^{\prime }}QL\rho _{S}\left(
t-t^{\prime }\right) \otimes \rho _{{\cal E}}\right) .
\]%
In summary the master equation reduces to 
\begin{equation}
\frac{\partial }{\partial t}\rho _{{\cal S}}\left( t\right) =L_{{\cal S}%
}^{eff}\left[ \rho _{{\cal S}}\left( t\right) \right] +\int_{0}^{t}{\cal K}%
\left( t^{\prime }\right) \left[ \rho _{S}\left( t-t^{\prime }\right) \right]
+{\cal J}\left( t\right)   \label{generalised master}
\end{equation}%
with 
\begin{eqnarray*}
L_{{\cal S}}^{eff} &\equiv &L_{S}+Tr_{{\cal E}}\left( L_{{\cal SE}}\rho _{%
{\cal E}}\right) , \\
{\cal K}\left( t\right) \left[ \rho _{S}\right]  &=&Tr_{{\cal E}}\left( L_{%
{\cal SE}}e^{QLt}QL\left[ \rho _{S}\otimes \rho _{{\cal E}}\right] \right) ,
\\
{\cal J}\left( t\right)  &=&Tr_{{\cal E}}\left( L_{{\cal SE}}e^{QLt}Q\rho
\left( 0\right) \right) .
\end{eqnarray*}%
In general it is an integro-differential equation with a memory kernel $%
{\cal K}$. Since the evolution of $\rho $ is unitary, the positivity of $%
\rho $ is maintained. The partial trace $\rho _{{\cal S}}\left( t\right) $
of the positive operator $\rho $ preserves the positivity. (\ref{generalised
master}) is exact and so guarantees a positive $\rho _{{\cal S}}\left(
t\right) $. It is only when approximations (truncations) are made that
positivity may be lost. The Markov approximation occurs if there is a
timescale $\tau _{{\cal E}}$ associated with ${\cal K}\left( t\right) $
which is \ much shorter than $\tau _{{\cal S}}$ the natural time scale of
the system ${\cal S}$ i.e. $\frac{\tau _{{\cal S}}}{\tau _{{\cal E}}}%
\longrightarrow \infty $. \ This Markov approximation has to be done
carefully for otherwise positivity can be lost\cite{davies}. Mathematically
there is another singular solution of this limit, $\tau _{{\cal S}%
}\longrightarrow \infty $ with $\tau _{{\cal E}}$ finite\cite{kossakowski}
which leads to the phenomenonology of dynamical semi-groups and the Lindblad
formalism \cite{lindblad}.

\bigskip

\begin{definition}
Time evolutions $\Lambda _{t}$ with $t\geq 0$ form a dynamical semi-group if
a) $\Lambda _{t_{1}}\circ \Lambda _{t_{2}}=\Lambda _{t_{1}+t_{2}}$, b) $Tr%
\left[ \Lambda _{t}\rho \right] =Tr\left[ \rho \right] $ for all $t$ and $%
\rho $ and c) are positive i.e map positive operators into positive
operators.
\end{definition}

\bigskip

There are other technical conditions such as strong continuity which we will
not dwell on. As far as applications are concerned the most important
characterisation of dynamical semi-groups is that they arise from the
singular limit mentioned above and are governed by the following theorem due
to Lindblad:

\bigskip

\begin{theorem}
If $P\left( {\frak H}\right) $ denotes the states on a Hilbert space ${\frak %
H}$, and $L$ is a bounded linear operator which is the generator of a
dynamical semi-group (i.e. $\Lambda _{t}=e^{Lt}$ ), then 
\[
L\left[ \rho \right] =-i\left[ H,\rho \right] +\frac{1}{2}\sum_{j}\left( %
\left[ V_{j}\rho ,V_{j}^{\dagger }\right] +\left[ V_{j},\rho V_{j}^{\dagger }%
\right] \right) 
\]%
where $H\left( =H^{\dagger }\right) ,\;V_{j}$ and $\sum_{j}V_{j}^{\dagger
}V_{j}$ are bounded linear operators on ${\frak H}$.
\end{theorem}

\bigskip This is the Lindblad form which has been used extensively in high
energy physics phenomenology. $L\left[ \rho \right] ,$ in the absence of the
terms involving the $V$s, is the Liouville operator. $H$ is the hamiltonian
which generally could be in the presence of a background stochastic
classical metric\cite{sarkar1} ( as we will discuss later). Such effects may
generally arise from back-reaction of matter within a quantum theory of
gravity \cite{hu} which decoheres the gravitational state to give a
stochastic ensemble description. In phenomenological analyses a theorem due to
Gorini, Kossakaowski and Sudarshan ~\cite{gorini} on the structure of $L$,
the generator of a quantum dynamical semi-group~\cite{lindblad,gorini} is of
importance. This states that for a non-negative matrix $c_{kl}$ (i.e. a matrix with
non-negative eigenvalues) such a generator is given by 
\[
\frac{{d\rho }}{{dt}}={\cal L}[\rho ]=-i[H,\rho ]+\frac{1}{2}%
\sum_{k,l}c_{kl}\left( [F_{k}\rho ,F_{l}^{\dagger }]+[F_{k},\rho
F_{l}^{\dagger }]\right) , 
\]%
where $H=H^{\dagger }$ is a hermitian Hamiltonian, $\{F_{k},k=0,...,n^{2}-1%
\} $ is a basis in $M_{n}({\bf C})$ such that $F_{0}=\frac{1}{\sqrt{n}}I_{n}$%
, Tr$(F_{k})=0~\forall k\neq 0$ and Tr$(F_{i}^{\dagger }F_{j})=\delta _{ij}$ %
\cite{gorini}. In applications we can take $F_{i}=\frac{{\Lambda _{i}}}{2}$
(where, for example, ${\Lambda _{i}}$ are the Gell-Mann matrices) and
satisfy the Lie algebra $[F_{i},F_{j}]=i\sum_{k}f_{ijk}F_{k},(i=1,...8),$ $%
f_{ijk}$ being the standard structure constants, antisymmetric in all
indices. It can always be arranged that the sum over $k$ and $l$ run over $%
1,\ldots ,8$. Without a microscopic model, in the three generation case, the
precise physical significance of the matrix $c_{kl}$ cannot be understood.
Moreover a general parametrisation of $c_{kl}$ is too complicated to have
any predictive power.

\bigskip

It is precise in formulation but gives no inkling of its ${\cal SE}$
compound system progenitor. Therein lies its weakness\cite{barenboim2} but
nonetheless it has been useful in providing `test' theories and estimating
orders of magnitudes for the strength of effects. If the strength of effects
are in accord with a theoretical picture then it has been customary to
conclude that the source of the decoherence is compatible with the
theoretical picture. Recently it has been argued that this may be too
simplistic and it is necessary to delve into the background ${\cal SE}$ to
be able to argue in favour of a picture.

\bigskip

\section{Master Equations from (Non-critical)\ String Theory}

\bigskip

When neutrinos from the Sun are produced ( e.g. from the nuclear $p-p$ cycle) and
pass through it, the nature of ${\cal E}$ and $L_{{\cal SE}}$ can be understood
from the gauge theories of the weak interactions\cite{burgess}. Consequently
the programme outlined in the previous paragraph with a perturbative
evaluation of ${\cal K}\left( t\right) $ is feasible in principle. However
in recent years there has been a debate on whether microscopic black holes
can induce quantum decoherence at a microscopic level. The presence of
quantum-fluctuating microscopic horizons, of radius of the order of Planck
length ($10^{-35}$ m), may give space-time a ``foamy'' structure, causing
decoherence of matter propagating in it. In particular, it has been
suggested~\cite{hawking} that such Planck-scale black holes and other
topological fluctuations in the space-time background cause a breakdown of
the conventional S-matrix description of asymptotic scattering in local
quantum field theory.Hence when we consider space-time foam we are on less
firm ground for applying the Lindblad formalism. Clearly gleaning an
understanding of the nature of space-time itself \ raises a huge number of
foundational issues. String theory is one attempt to address such questions
but is still far from the goal of clarifying strong gravity. There are some
who even believe that gravity is an emergent feature and consequently that
an attempt to understand the quantum aspects of gravity may be fundamentally
futile. It is not appropriate to enter this debate here. As far as
experiments are concerned, both now and in the near future, it is reasonable
to ask what the current theories have to say concerning quantum effects
where a nearly flat metric gravity is clearly reasonable.

\bigskip

The issue of quantum-gravity-induced decoherence is controversial and worthy
of further phenomenological exploitation. We shall restrict ourselves to a
specific framework for analyzing decoherent propagation of low-energy matter
in foamy space-time backgrounds in the context of string theory~\cite%
{strings,polch}, the so-called Liouville-string~\cite{ddk} decoherence~\cite%
{emn}. One motivation for using string theory is that it appears 
to be the best controlled theory of quantum gravity available to date. At
this juncture we should also mention that there are other interesting
approaches to quantum space-time foam, which also lead to experimental
predictions, e.g. the ``thermal bath'' approach advocated in \cite{garay},
according to which the foamy gravitational environment may behave as a
thermal bath; this induces decoherence and diffusion in the propagating
matter, as well as quantum damping in the evolution of low-energy
observables, features which are, at least in principle, testable
experimentally. As we shall see presently, similar behaviour is exhibited by
the specific models of foam that we study here; the D-particle foam model of %
\cite{emw,recoil} may characterize modern versions of string theory~\cite%
{polch}, and are based on point-like membrane defects in space-time
(D-particles). Such considerations have more recently again come to the fore
because of current neutrino data including LSND data~\cite{reviewnu}. There
is experimental evidence, that the neutrino has mass which leads to neutrino
oscillations. However LSND results appear consistent with the dominance of
anti-neutrino oscillations $\overline{\nu }_{e}\rightleftarrows $ $\overline{%
\nu }_{\mu }$ over neutrino oscillations. In particular, provided LSND
results turn out to be correct, which at present is quite unclear, there is
evidence for CPT violation. It has been suggested recently~\cite{barenboim}
that Planck scale quantum decoherence may be a relevant contribution to the
CPT violation seen in the experiments of LSND. Other examples of flavour
oscillating systems with quite different mass scales are furnished by $B%
\overline{B}$ and $K\overline{K}$ systems~\cite{kaons}. The former because
of the large masses involved provides a particularly sensitive system for
investigating the Planck scale fluctuations embodied by space-time foam. In
all these cases, experiments, such as CPLEAR~\cite{cplear}, provide very low
bounds on CPT violation which are not inconsistent with estimates from
dimensional analysis for the magnitudes of effects from space-time foam.
These systems have been analyzed within a dynamical semigroup approach to
quantum Markov processes. Once the framework has been accepted then a master
equation for finite-dimensional systems ensued which was characterized by a
small set of parameters. This approach is somewhat phenomenological and is
primarily used to fit data \cite{ehns, lisi, benatti}. Consequently it is
important to obtain a better understanding of the nature of decoherence from
a more fundamental viewpoint.

\bigskip

Given the very limited understanding of gravity at the quantum level, the
analysis of modifications of the quantum Liouville equation implied by
non-critical strings can only be approximate and should be regarded as
circumstantial evidence in favour of the dissipative master equation. In the
context of two-dimensional toy black holes\cite{2dbhstring} and in the
presence of singular space-time fluctuations there are believed to be
inherently unobservable delocalised modes which fail to decouple from light
(the observed) states. The effective theory of the light states which are
measured by local scattering experiments can be described by a non-critical
Liouville string. This results in an irreversible temporal evolution in
target space with decoherence and associated entropy production.

\bigskip

The following master equation for the evolution of stringy low-energy matter
in a non-conformal $\sigma $-model~can be derived\cite{emn} 
\begin{equation}
\partial _{t}\rho =i\left[ \rho ,H\right] +:\beta ^{i}{\cal G}_{ij}\left[
g^{j},\rho \right] :  \label{master}
\end{equation}%
where $t$ denotes time (Liouville zero mode), the $H$ is the effective
low-energy matter Hamiltonian, $g^{i}$ are the quantum background target
space fields, $\beta ^{i}$ are the corresponding renormalization group $%
\beta $ functions for scaling under Liouville dressings and ${\cal G}_{ij}$
is the Zamolodchikov metric \cite{zam,kutasov} in the moduli space of the
string. The double colon symbol in (\ref{master}) represents the operator
ordering $:AB:=\left[ A,B\right] $ . The index $i$ labels the different
background fields as well as space-time. Hence the summation over $i,j$ in (%
\ref{master}) corresponds to a discrete summation as well as a covariant
integration $\int d^{D+1}y\,\sqrt{-g}$\bigskip\ where $y$ denotes a set of $%
\left( D+1\right) $-dimensional target space-time co-ordinates and $D$ is
the space-time dimensionality of the original non-critical string.

\bigskip

\begin{figure}[htb]
\begin{center}
\epsfxsize=2in
\centerline{\epsffile{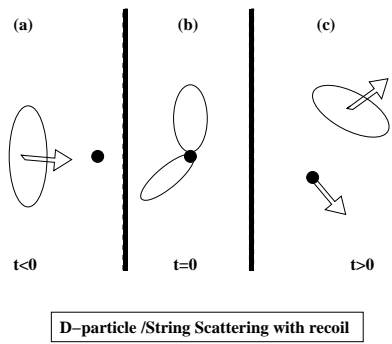}}
\caption{{\it Schematic picture of the scattering of a string matter state
on a D-particle, including recoil of the latter. The sudden impulse at $t=0$%
, implies a back reaction onto the space time, which is described by a
logarithmic conformal field theory. The method allows for the perturbative
calculation of the induced space-time distortion due to the entangled state
in (b).}}
\label{fig:dfoam}
\end{center}
\end{figure}

The discovery of new solitonic structures in superstring theory~\cite{polch}
has dramatically changed the understanding of target space structure. These
new non-perturbative objects are known as D-branes and their inclusion leads
to a scattering picture of space-time fluctuations. Heuristically, when low
energy matter given by a closed (or open) string propagating in a $\left(
D+1\right) $-dimensional space-time collides with a very massive D-particle
embedded in this space-time, the D-particle recoils as a result. Since there
are no rigid bodies in general relativity the recoil fluctuations of the
brane and their effectively stochastic back-reaction on space-time cannot be
neglected. On the brane there are closed and open strings propagating. Each
time these strings cross with a D-particle, there is a possibility of being
attached to it, as indicated in Fig. \ref{fig:dfoam}. The entangled state
causes a back reaction onto the space-time, which can be calculated
perturbatively using logarithmic conformal field theory formalism~\cite{kmw}%
. \ 

Now for large Minkowski time $t$, the non-trivial changes from the flat metric produced
from D-particle collisions are

\begin{equation}
g_{0i}\simeq \overline{u}_{i}\equiv \frac{u_{i}}{\varepsilon }\propto \frac{%
\Delta p_{i}}{M_{P}}  \label{recoil}
\end{equation}
where $u_i$ is the velocity and $\Delta p_{i}$ is the momentum transfer during a collision, $\varepsilon ^{-2}$ is identified with $t$ and $M_{P}$
is the Planck mass (actually, to be more precise $M_{P}=M_{s}/g_{s}$, where $%
g_{s}<1$ is the (weak) string coupling, and $M_{s}$ is a string mass scale);
so $g_{0i}$ is constant in space-time but depends on the energy content of
the low energy particle and the Ricci tensor $R_{MN}=0$ where $M$ and $N$
are target space-time indices. Since we are interested in fluctuations of
the metric the indices $i$ will correspond to the pair $M,N$. However,
recent astrophysical observations from different experiments all seem to
indicate that $73$\% of the energy of the Universe is in the form of dark
energy. Best fit models give the positive cosmological constant
Einstein-Friedman Universe as a good candidate to explain these
observations. For such de Sitter backgrounds $R_{MN}\propto \Omega g_{MN}$
with $\Omega >0$ a cosmological constant. Also in a perturbative derivative
expansion (in powers of $\alpha ^{\prime }$ where $\alpha ^{\prime
}=l_{s}^{2}$ is the Regge slope of the string and $l_{s}$ is the fundamental
string length) in leading order 
\begin{equation}
\beta _{\mu \nu }=\alpha ^{\prime }R_{\mu \nu }=\alpha ^{\prime }\Omega
g_{\mu \nu }  \label{fixedpoint}
\end{equation}%
and 
\begin{equation}
{\cal G}_{ij}=\delta _{ij}.
\end{equation}%
This leads to 
\begin{equation}
\partial _{t}\rho =i\left[ \rho ,H\right] +\,\alpha ^{\prime }\Omega :g_{MN}%
\left[ g^{MN},\rho \right] :  \label{master2}
\end{equation}%
For a weak-graviton expansion about flat space-time, $g_{MN}=\eta
_{MN}+h_{MN}$, and 
\begin{equation}
h_{0i}\propto \frac{\Delta p_{i}}{M_{P}}.  \label{recoil2}
\end{equation}%
If an antisymmetric ordering prescription is used, then the master equation
for low energy string matter assumes the form%
\begin{equation}
\stackrel{.}{\partial _{t}\rho _{Matter}}=i\left[ \rho _{Matter},H\right]
-\,\Omega \left[ h_{0j},\left[ h^{0j},\rho _{Matter}\right] \right]
\label{master3}
\end{equation}%
( when $\alpha ^{\prime }$ is absorbed into $\Omega )$. In view of the
previous discussion this can be rewritten as%
\begin{equation}
\stackrel{.}{\partial _{t}\rho _{Matter}}=i\left[ \rho _{Matter},H\right]
-\,\Omega \left[ \overline{u}_{j},\left[ \overline{u}^{j},\rho _{Matter}%
\right] \right] ~.  \label{master4}
\end{equation}%
thereby giving the {\it master equation for Liouville decoherence} in the
model of a D-particle foam with a cosmological constant.

\bigskip

The above D-particle inspired approach deals with possible non-perturbative
quantum effects of gravitational degrees of freedom. The analysis is
distinct from the phenomenology of dynamical semigroups which does not
embody specific properties of gravity. Indeed the phenomenology is
sufficiently generic that other mechanisms of decoherence such as the MSW
effect can be incorporated within the same framework. Consequently an
analysis which is less generic and is related to the specific \ decoherence
implied by non-critical strings is necessary.It is sufficient to study a
massive non-relativistic particle propagating in one dimension to establish
qualitative features of D-particle decoherence. The environment will be
taken to consist of both gravitational and non-gravitational degrees of
freedom; hence we will consider a generalisation of quantum Brownian motion
for a particle which has additional interactions with D-particles. This will
allow us to compare qualitatively the decoherence due to different
environments.The non-gravitational degrees of freedom in the environment (in
a thermal state) are conventionally modelled by a collection of harmonic
oscillators with masses $m_{n}$, frequency $\omega _{n}$ and co-ordinate
operator $\widehat{q}_{n}$ coupled to the particle co-ordinate $\widehat{x}$
by an interaction of the form $\sum_{n}g_{n}\widehat{x}\widehat{q}_{n}$. The
master equation which is derived can have time dependent coefficients due to
the competing timescales, e.g. relaxation rate due to coupling to the
thermal bath, the ratio of the time scale of the harmonic oscillator to the
thermal time scale etc. However an ab initio calculation of the
time-dependence is difficult to do in a rigorous manner. It is customary to
characterise the non-gravitational environment by means of its spectral
density $I\left( \omega \right) \left( =\sum_{n}\delta \left( \omega -\omega
_{n}\right) \frac{g_{n}^{2}}{2m_{n}\omega _{n}}\right) $. The existence of
the different time scales leads in general to non-trivial time dependences
in the coefficients in the master equation which are difficult to calculate
in a rigorous manner \cite{BLH1992}. The dissipative term in (\ref{master4})
involves the momentum transfer operator due to recoil of the particle from
collisions with D-particles (\ref{recoil}). This transfer process will be
modelled by a classical Gaussian random variable $r$ which multiplies the
momentum operator $\widehat{p}$ for the particle: 
\begin{equation}
\overline{u_{i}}\qquad \rightarrow \qquad \frac{r}{M_{P}}\widehat{p}
\label{trnsf}
\end{equation}%
Moreover the mean and variance of $r$ are given by 
\begin{equation}
\left\langle r\right\rangle =0~,\qquad {\rm and}\qquad \left\langle
r^{2}\right\rangle =\sigma ^{2}~.  \label{random}
\end{equation}%
On amalgamating the effects of the thermal and D-particle environments, we
have for the reduced master equation \cite{sarkar} for the matter (particle)
density matrix $\rho $ (on dropping the Matter index)%
\begin{equation}
i\frac{\partial }{\partial t}\rho =\frac{1}{2m}\left[ \widehat{p}^{2},\rho %
\right] -i\Lambda \left[ \widehat{x},\left[ \widehat{x},\rho \right] \right]
+\frac{\gamma }{2}\left[ \widehat{x},\left\{ \widehat{p},\rho \right\} %
\right] -i\Omega r^{2}\left[ \widehat{p},\left[ \widehat{p},\rho \right] %
\right]  \label{master5}
\end{equation}%
where $\Lambda ,\gamma $ and $\Omega $ are real time-dependent coefficients.
As discussed in \cite{sarkar} a possible model for $\Omega \left( t\right) $
is 
\begin{equation}
\Omega \left( t\right) =\Omega _{0}+\frac{\widetilde{\gamma }}{a+t}+\frac{%
\widetilde{\Gamma }}{1+bt^{2}}  \label{cosmological}
\end{equation}%
where $\omega _{0}$, $\widetilde{\gamma }$, $a$, $\widetilde{\Gamma }$ and $%
b $ are positive constants. The quantity $\widetilde{\gamma }<1$ contains
information on the density of D-particle defects on a four-dimensional
world.The time dependence of $\gamma $ and $\Lambda $ can be calculated in
the weak coupling limit for general $n$ (i.e. ohmic, $n=1$ and non-ohmic $%
n\neq 1$ environments)$\ $where 
\begin{equation}
I\left( \omega \right) =\frac{2}{\pi }m\gamma _{0}\omega \left[ \frac{\omega 
}{\varpi }\right] ^{n-1}e^{-\omega ^{2}/\varpi ^{2}}  \label{spectral}
\end{equation}%
and $\varpi $ is a cut-off frequency. The precise time dependence is
governed by $\Lambda \left( t\right) =\int_{0}^{t}ds\,\nu \left( s\right) $
and $\gamma \left( t\right) =\int_{0}^{t}ds\,\nu \left( s\right) s$ where $%
\nu \left( s\right) =\int_{0}^{\infty }d\omega \,I\left( \omega \right)
\coth \left( \beta \hbar \omega /2\right) \cos \left( \omega s\right) $. For
the ohmic case, in the limit $\hbar \varpi \ll k_{B}T$ followed by $\varpi
\rightarrow \infty $, $\Lambda $ and $\gamma $ are given by $m\gamma
_{0}k_{B}T$ and $\gamma _{0}$ respectively after a rapid initial transient.
For high temperatures $\Lambda $ and $\gamma $ have a powerlaw increase with 
$t$ for the subohmic case whereas there is a rapid decrease in the
supraohmic case.

\bigskip

\section{CPT and Recoil}

\bigskip

The above model of space-time foam refers to a specific string-inspired
construction. However the form of the induced back reaction (\ref{recoil})
onto the space-time has some generic features, and can be understood more
generally in the context of effective theories of such models, which allows
one to go beyond a specific non-critical (Liouville) model. Indeed, the
D-particle defect can be viewed as an idealisation of some (virtual,
quantum) black hole defect of the ground state of quantum gravity, viewed as
a membrane wrapped around some small extra dimensions of the (stringy) space
time, and thus appearing to a four-dimensional observer as an
``effectively'' point like defect. The back reaction on space-time due to
the interaction of a pair of neutral mesons, such as those produced in a
meson factory, with such defects can be studied generically as follows:
consider the non-relativistic recoil motion of the heavy defect, whose
coordinates in space-time,in the laboratory frame, are $%
y^{i}=y_{0}^{i}+u^{i}t$, with $u^{i}$ the (small) recoil velocity.One can
then perform a (infinitesimal) general coordinate transformation $y^{\mu
}\rightarrow x^{\mu }+\xi ^{\nu }$ so as to go to the rest (or co-moving )
frame of the defect after the scattering. From a passive point of view, for
one of the mesons, this corresponds to an induced change in metric of
space-time of the form (in the usual notation, where the parenthesis in
indices denote symmetrisation) $\delta g_{\mu \nu }=\partial _{(\mu }\xi
_{\nu )}$, which in the specific case of non-relativistic defect motion
yields the off-diagonal metric elements (\ref{recoil}). Such transformations
cannot be performed simultaneously for both mesons, and moreover in a full
theory of quantum gravity the recoil velocities fluctuate randomly, as we
shall discuss later on. This means that the effects of the recoil of the
space-time defect are observable. The mesons will feel such effects in the
form of induced fluctuating metrics (\ref{recoil}). It is crucial to note
that the interaction of the matter particle (meson) with the foam defect may
also result in a ``flavour'' change of the particle (e.g. the change of a
neutral meson to its antiparticle). This feature can be understood in a
D-particle Liouville model by noting that the scattering of the matter probe
off the defect involves first a splitting of a closed string representing
matter into two open ones, but with their ends attached to the D-particle,
and then a joining of the string ends in order to re-emit a closed string
matter state. The re-emitted (scattered) state may in general be
characterised by phase, flavour and other quantum charges which may not be
required to be conserved during black hole evaporation and disparate
space-time-foam processes. In our application we shall restrict ourselves
only to effects that lead to flavour changes. The modified form of the
metric fluctuations (\ref{recoil}) of each component of the metric tensor $%
g^{\alpha \beta }$ will not be simply given by the simple recoil distortion (%
\ref{recoil}), but instead can be taken to have a $2\times 2$ (``flavour'')
structure \cite{bernabeu}: 
\begin{align}
g^{00}& =\left( -1+r_{4}\right) {\sf 1}  \nonumber \\
g^{01}& =g^{10}=r_{0}{\sf 1}+r_{1}\sigma _{1}+r_{2}\sigma _{2}+r_{3}\sigma
_{3}  \label{metric} \\
g^{11}& =\left( 1+r_{5}\right) {\sf 1}  \nonumber
\end{align}%
where ${\sf 1}$ , is the identity and $\sigma _{i}$ are \ the Pauli
matrices. The above parametrisation has been taken for simplicity and we
can also consider motion to be in the $x$- direction which is natural since
the meson pairs move collinearly. A metric with this type of structure is
compatible with the view that the D-particle defect is a ``point-like''
approximation for a compactified higher-dimensional brany black hole, whose
no hair theorems permit non-conservation of flavour.(In the case of neutral
mesons the concept of ``flavour'' refers to either particle/antiparticle
species or the two mass eigenstates). The detailed application of this model
to the $\omega $ effect for neutral mesons can be found in \cite{bernabeu}.

\bigskip\ 

\bigskip

\begin{acknowledgement}
I would like to thank A. di Domenico for the invitation to contribute to the Daphne Physics Handbook. Discussions with N E\ Mavromatos, J Bernabeu and A\ Waldron-Lauda are gratefully acknowledged.
\end{acknowledgement}

\end{document}